# Electromagnetic Radiation from Double Layers

N. Brenning*, M. E. Koepke**, I. Axnäs*, and M. A. Raadu*

*Alfvén Laboratory, Royal Institute of Technology, se-100 44 Stockholm, Sweden
**Department of Physics, West Virginia University, Morgantown, USA

## Abstract

The background motivation, and some preliminary results, are reported for a recently begun investigation of a potentially important mechanism for electromagnetic radiation from space, Double Layer Radiation (DL-radiation). This type of radiation is proposed to come from DL-associated, spatially localized high frequency (hf) spikes that are driven by the electron beam on the high-potential side of the double layer. It is known, but only qualitatively, from laboratory experiments that double layers radiate in the electromagnetic spectrum. In our experiment the spectrum has high amplitude close to the plasma frequency, several hundred MHz. No clear theoretical model exists today. The quantitative evaluation is complicated because measurements are made in the near field of the radiating structure, and in the vicinity of conducting laboratory hardware that distorts the field. Because the localized electrostatic wavelengths (approximately 1 cm) can be relatively small compared to the emitted electromagnetic wavelengths, the situation is further complicated. We discuss the mutual influence between the ion density profile and hf-spike formation, and propose that some kind of self-organization of the density profile through the ponderomotive force of the hf spike might be operating. First results regarding the electromagnetic radiation are reported: the frequency, the time variation of the amplitude, and the spatial distribution in the discharge vessel.

## 1. Introduction.

What information we obtain from space is almost exclusively reaching us in the form of electromagnetic radiation. More than 99% of it comes from matter in the plasma state. Plasmas in space are usually in relative motion and electromagnetic radiation often comes from regions where plasmas with different velocities interact. The proposed models often involve some kind of electric dynamo process leading to acceleration of charged particles that then radiate, either as single particles by synchrotron radiation or bremsstrahlung, or by some collective mechanism. Examples are many; here are just a few examples.

- Type III solar bursts in the solar wind [1], emitted from the solar corona, which are associated with electrons accelerated in flares or active regions.
- The auroral kilometric radiation [2] that is ultimately driven by the solar wind interaction with the Earth's magnetosphere.
- The Jovian radio emission [3], which is associated with the current system driven between the moon Io and Jupiter's co-rotating magnetosphere.
- The transient radio emission [4], driven by the repeated passage of a pulsar (with associated pulsar wind) through the circumstellar disc of a Be star.
- Pulsed radiation from neutron star winds [5], driven by the currents (*i. e.,* changes in magnetic field orientation) which are frozen into the wind from an off-centre rotating stellar magnetosphere.

Models are proposed for several of these processes. However, it is a lesson from the exploration of the near space plasma that models built on incomplete knowledge can easily go astray. Elaborate and widely believed descriptions of the plasma processes in the Earth's magnetosphere had to be completely rewritten when satellite measurements began to give reliable in-situ data.

Mechanims for plasmas to produce electromagnetic radiation can be very complicated (nonlinear, with complicated geometry and needing full electromagnetic modelling), and therefore unlikely to be discovered from purely theoretical analysis. A complete inventory of all possible mechanisms for plasmas in relative motions to radiate is therefore important. Double Layer Radiation from DL-associated hf spikes is one such mechanism, with potentially large efficiency, and with properties that are today not known.

## 2. Double Layer Radiation

Double layers are current-carrying structures where very strong local electric field can be sustained within a limited region of space. It is known from laboratory experiments that double layers arise in a plasma at sufficiently high current densities. Recent observations with the FAST satellite have shown that they can occur naturally in space [6]. A double layer can concentrate the release of electric energy both in space and in time. The spatial compression is because the electric energy delivered from an extended electric circuit is dissipated (transferred to particle energy) within a small region, with an extent of typically some Debye lengths. A large temporal compression is also possible because a double layer does not behave as a usual linear circuit element. Once it starts to build up, it has a negative differential resistance, which can lead to explosive double layers where magnetic energy that has been built up and stored in a circuit for a long time is released during a short time [7].

It is known, but only qualitatively, from laboratory experiments that double layers radiate in the electromagnetic spectrum. In the 1990's, both Martin Volwerk [8] and Lennart Lindberg [9] made studies in the laboratory, using magnetic pickup coils to measure the electromagnetic radiation from double layers in a mercury plasma machine. The spectrum was found to contain characteristic peaks as the electron gyro- and plasma frequencies, and combinations of them, but also other frequencies that might be apparatus-dependent. No clear theoretical model emerged from these investigations, and no absolute calibrated values of the radiation strength could be obtained. These measurements were made in the near field of the radiating structure, and in a laboratory device with conducting parts that distort the field and make quantitative evaluations very difficult. The plasma parameters and length scales for the experiments lead to a complicated situation, where the internal wavelengths may be relatively small but where the external electromagnetic field has relatively long wavelengths (*e.g.* 50 cm at 600 MHz, a typical plasma frequency). It was suggested by Jan Kuijpers [10] that the radiation might arise directly from the acceleration that single charged particles experience in the double layer, but the theory was not elaborated to match the experimental observations.

The first space application to our knowledge of DL-radiation is a proposal by Borovsky [11] regarding radiation from the arms of double radio galaxies, which can carry currents which have been estimated to be as large as $10^{17}$ A. The radiation mechanism was only schematically described as follows. The electron beams produced by a double layer, maintained by this large current, should drive two-stream instabilities in the ambient plasma on the high potential side. Bunched electrons in the instability's wave structures should then act "in the fashion of a free electron laser" and produce electromagnetic radiation.

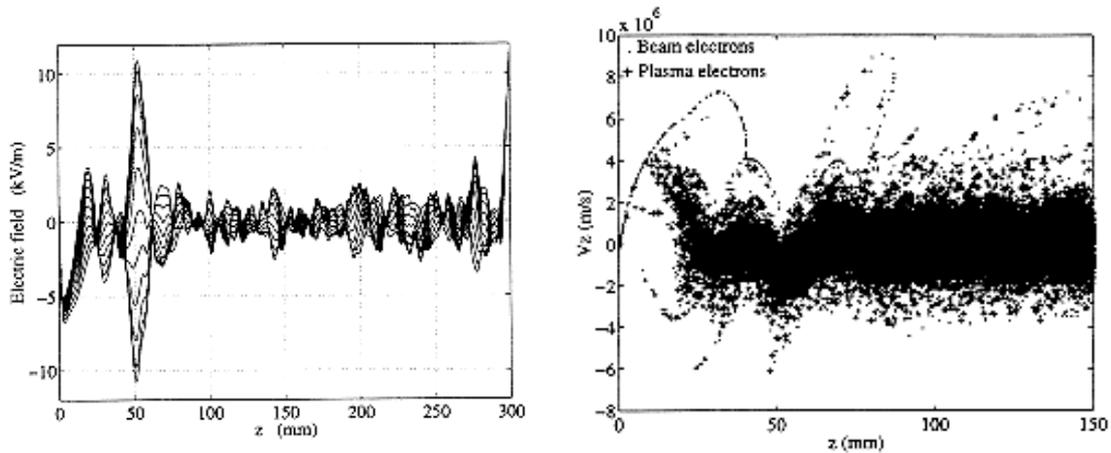

Fig. 1. Results from a 1-D particle-in-cell simulation of an experimentally created hf spike. Left figure: the electric field plotted at 11 different instants of time during one half period of the oscillation. The hf spike is centred at $z = 50$ mm. Right hand figure: a phase space diagram for the electrons, showing that the beam of electrons coming from the left is completely scattered at the hf spike, here centred at $z = 45$ mm. From [12].

## 3. DL-radiation from a high frequency (hf-) spike.

A step towards a better understanding was taken at our laboratory in the mid-90's. We started a study of the instabilities driven by the electron beam on the high potential side of the double layer. The work resulted in three PhD theses: Gunell 1997 [13], Löfgren 1999 [14], and Wendt 2001 [15]. New probes techniques were developed [16] to enable reliable absolute calibrated electric field measurements at high frequencies (several 100 Mhz). The laboratory results were interpreted with the help of particle computer simulations and analytical modelling [17, 18, 19]. We found that the oscillations become concentrated to very narrow, high amplitude wave features. Characteristic features of these "hf spikes" are:

• width: (half-amplitude width) less than one wavelength
• frequency: the local plasma frequency
• velocity of the envelope: the ion acoustic velocity
• phase velocity: the electron beam velocity
• amplitude: up to 10 kV/m (Fig. 1, left).

The local wave amplitude in the spike is high enough to completely scatter the electron beam in one transit (Fig 1, right). The ponderomotive force of the wave changes the ion density on the ion acoustic time scale (Fig. 2). We regard it as very likely that such strong nonlinear structures, with violent accelerations of bunches of electrons over short distances, and with strong gradients in amplitude both along the beam direction and across it, can be very efficient in coupling to electromagnetic radiation.

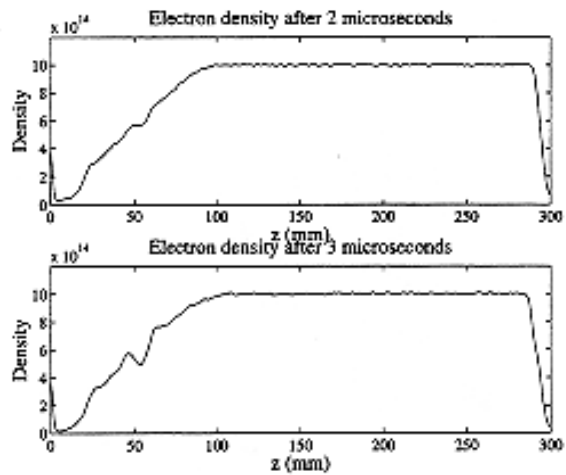

Fig. 2. The plasma density after 2 and 3 μs, respectively. A density dip is observed to develop at the position of the hf spike, at 50 mm (See Fig. 1, left panel). From [12].

The proposal that strong DL-radiation might be produced not only in the laboratory but also in space is so far built on a combination of known facts and extrapolations:

- Double layers can be created quite easily, in a wide variety of apparatus configurations and parameter regimes. It only requires that some critical limit in electric current density be exceeded. Double layers have also recently been observed with certainty for the first time in space, in the auroral current system.

- A double layer taps the electric energy from a large region of space and delivers it into a limited volume, and sometimes also concentrate the energy release in time [7]. The major part of the energy appears in the form of electron beams in the high potential plasma.

- In every device in the laboratory where we have looked for them so far (four different experiments in total), the electron beams create high-amplitude hf spikes of small spatial extent. There is the possibility that there is some self-organizing mechanism to achieve this.

- In one of our devices the hf spike structures have been extensively studied and modeled by 1-D computer simulations. The results lead us to believe that they can be very efficient in coupling to electromagnetic radiation (by indirect reasoning; this cannot be adequately modeled one dimension).

Major things, however, remain to be done if reliable extrapolations to space shall be made. The basic reason why hf spikes appear regularly in so many different devices is not identified. We do know, however, from the simulations that the precise plasma density profile on the high density side is crucial. It is known from experiments that double layers do radiate electromagnetically, but the radiating efficiency is not known. Regarding the physical understanding of the radiating mechanism we have only general ideas.

# 4. Experimental results

## 4.1. The statistical occurrence of hf spikes at a double layer: self organization?

In laboratory experiments in various geometries, a hf spike seems to arise almost inevitably in connection with a double layer. However, there are contradicting (unpublished) indications from the 1-D computer simulations of ref [17, 18]: The simulation can be started many times with slightly different approximations of the experimental ion density profile, but without producing a hf spike. The beam-plasma interaction is instead spread out over a wide region of the high potential plasma. Only after a lot of trial and error is there a run in which a hf spike arises, which then shows all the detailed characteristics of those seen in the laboratory. There has been no systematic test of how statistically probable a hf spike is in the simulations, but fingertip feeling puts it somewhere around 10%, perhaps much lower - in spite of the fact that the ion density profiles always are chosen close to those that eventually succeed.

The ion density profile seems to determine if a hf spike arises or not. In a simulation where a hf spike has grown up (which can take some microseconds, many tens of electron transit times), an interesting test can be made. The fully grown hf spike involves strong perturbations of the electron beam (See fig. 1, the right hand panel) but only marginal motion of the heavy ions. If now all these electrons are replaced with thermal background electrons, plus a cold beam from the cathode, the hf spike re-appears practically immediately, on the electron transit time scale. The ion profile is completely decisive for the occurrence of a hf spike.

The question arises if there is a mechanism in the experiments that maintain a hf-spike-producing ion density profile. One hypothesis is that there is some self-organization by the ponderomotive force. A seen in Fig. 2, the ponderomotive force in the hf spike is large enough to push about the ions on the microsecond time scale. The presence of a hf spike therefore influences the ion density profile. If this acts in a direction such that the *disappearance* of the hf spike would make the ion density profile relax towards a profile which is *more favourable* for hf spike formation, then this would close a Self-Organized Criticality (SOC) type of feedback loop where the hf spike and the ion density profile interact constructively.

## 4.2. Preliminary experiments on the radiation from a hf spike

A series of preliminary experiments have been made in a plasma device (the Green Tank) with a mirror magnetic field configuration, and with an electron beam emitted from the mirror into the plasma from a heated lantanium hexaboride cathode. The configuration is the same as used in [12,13]. It produces a stationary cathode sheath that we here use instead of a free double layer in order to study, under more controlled conditions, the hf spike formation and the radiation from it. We have used magnetic pickup coils to detect the electromagnetic radiation. The azimuthal component of the magnetic disturbance is strongest, and only this component is used for the preliminary investigations reported below.

*Antenna action of hf probes in the spike.*
Ideally one would like to measure the electrostatic hf wave oscillations inside the hf spike (the es-oscillations), and simultaneously measure the electromagnetic radiation in the surrounding space (the em-radiation). Such measurements, however, become corrupted by the fact that the hf probes, when inserted into the spike, act as antennas: the amplitude of the surrounding em-radiation increases by typically a factor 2-4 by the presence of a hf probe in the hf spike region. We therefore have measured the em-radiation without hf probes in the spike. Only comparisons of indirect, or statistical, nature can therefore (so far) be made between the es-oscillations and the em-radiation.

Since the presence of hf probes in the hf spike change the em-radiation amplitude, the question arises if they only act as antennas or if their presence influences the hf spike itself. We have come to the conclusion that this is not the case: they act as passive antennas, driven by the hf spike oscillations. This conclusion is based on two observations. First, a series of tests has shown that the hf spike features, as obtained by a single probe, are not significantly influenced by the presence of another probe that is moved around. Second, computer simulations (which have no counterpart of a disturbing probe in them) reproduce the probe data on the hf spike quite well [13].

*Amplitude variations on the microsecond (ion acoustic) time scale*
The time variations of the em-radiation show two characteristic time scales: nanoseconds for the oscillations themselves, and microseconds for the amplitude variations. Long time series recordings, with random trigger, show that the em-radiation is seldom completely absent, but varies significantly in amplitude. The typical time scale for amplitude variations is microseconds, and the radiation usually has a bursty nature. All these features, the frequency, the long time statistical occurrence, and the μs-time scale amplitude variations, also characterise the es-oscillations inside the hf spike. This supports our opinion that the em-radiation originates there. The microsecond scale variations might be associated with perturbations of the ion density profile, on the ion acoustic time scale, as demonstrated in the simulations, see Fig. 2.

The evaluation of the em-radiation signals is complicated by the fact that there is a spatial variation involved: the em amplitude maxima do sometimes, but not always, occur simultaneously at different locations in the plasma chamber. There can therefore not be a one-to-one correlation between the es wave amplitude in the hf spike and the em-radiation everywhere. Multi-point vector-resolved measurements of the em-radiation might be required for understanding.

*Herring bone patterns: signs of mode competition?*
An intriguing feature is that there are frequently trains of bursts in the em-radiation, in the form of herringbone patters. These patterns could either be due to beating between two closely lying frequencies, or be signs of mode transfer. The latter interpretation is supported by observations of jumps in frequency, and of frequency modulations at burst-burst junctions.

*Frequency variations with plasma parameters (B, $n_e$).*
The waveforms in a small time window generally looks quite monochromatic. Spectra taken over a longer time are broader, and typically show a broad maximum somewhere in the range 300 – 500MHz, in the range somewhat above the plasma frequency at the position of the hf spike in the discharge. With a frequency analyzer we have inspected the spectra visually, while changing experimental parameters in the plasma device, by varying the mirror magnetic field strength and the discharge current. The preliminary conclusion is that the frequency is associated with the plasma frequency rather than with the electron gyro frequency.

*Screening of hf em radiation from the plasma core?*
The spatial distribution of the hf (300-500MHz) em radiation has been mapped in the plasma chamber. It was found to fill the whole chamber, except a central rotationally symmetric region, anchored in the volume where the hf spike is located, and from there mapped out into the plasma chamber along the (diverging) magnetic field lines. Our preliminary interpretation is that this region has such a high plasma density than the em radiation is cut off. This agrees with the earlier findings (see Fig. 2) that the hf spike is located at a position where the plasma density is lower than in the central plasma. If the hf spike radiates in all directions, with close to the local plasma frequency, the radiation would be excluded from the main plasma as observed.

*Unidentified lf radiation (20-40MHz) in the plasma core*
In the central plasma, some high frequency (300-500MHz) remains with much lower amplitude, but there is instead a high amplitude lower frequency (20-40 MHz) signal, which is not yet investigated in any detail. These oscillations are absent outside the plasma core. They are picked up by the magnetic probe, but they cannot be purely electromagnetic: they would be evanescent in the local high plasma density. They are in a frequency range between the ion and electron gyro frequencies. It is interesting to speculate that they might be whistler waves; like electromagnetic radiation, whistlers can be detected in space far from the source, and were actually proposed to be excited in double layers already by Volwerk [7].

## 5. Summary.

Three steps are necessary for the electromagnetic double-layer radiation mechanism we want to study. First, a double layer must be created in a current-carrying plasma. Second, the electron beam created on the high potential side of the double layer must create a localized hf spike. Third, the hf spike must couple to electromagnetic waves and act as a sender-antenna. If all three are fulfilled, there is a possibility of very high efficiency in converting electric energy to radiation.

The first of these two steps have been studied earlier in experiments and models. It has been known for a long time that double layers can be created quite easily, in a wide variety of apparatus configurations and parameter regimes. They basically require that some critical limit in electric current density be exceeded. Double layers have also recently been observed with certainty for the first time in space, in the auroral current system. Experiments during the last decade have furthermore shown that hf spikes generally are formed on the high potential side of double layers.

This work reports some first results regarding studies of the last step, the production of electromagnetic radiation from a hf-spike. Preliminary conclusions are that there is indeed em-radiation most of the time, that it is dominated by frequencies around the local plasma frequency, and that it is bursty and irregular in time on the ion acoustic time scale. This electromagnetic radiation however does not penetrate the high density plasma on the high potential side of the double layer. In that part or the plasma, there is instead a lower frequency wave excited, which might be a whistler wave. Future work will be directed towards understanding of the excitation mechanisms, and in quantifying the efficiency in converting the electric energy deposited in the double layer to radiation.

## Acknowledgements

This work was supported by the Swedish National Space Board.